\documentclass[a4paper,11pt]{article}
\usepackage{pos}

\usepackage{bm}
\usepackage{epsfig}
\usepackage{graphicx}
\usepackage[export]{adjustbox}
\usepackage{amsmath}

\usepackage{amssymb}
\usepackage{mathrsfs}
\usepackage{color}
\usepackage[usenames,dvipsnames]{xcolor}
\usepackage{hyperref}
\usepackage{feynmp-auto}
\usepackage{tikz}
\usepackage{braket}
\usepackage[caption=false]{subfig}
\usepackage{multirow}
\usepackage{array}
\usepackage{slashed}
\usepackage{rotating}
\usepackage{caption}
\usepackage{subcaption}
\usepackage{bbold}

\newcommand{\nb}{\bar{n}}
\newcommand{\LQCD}{\Lambda_{\rm QCD}}
\newcommand{\xeff}{x_{\rm eff}}
\newcommand{\Aeff}{A_{\rm eff}}

\title{Resummed Power Corrections in Nuclear DVCS}

\author*[a]{John Terry}

\affiliation[a]{Physics Division, Argonne National Laboratory,\\ Lemont, IL 60439, USA}

\emailAdd{terryj@anl.gov}

\abstract{%
Generalized Parton Distributions (GPDs) encode the three-dimensional structure
of hadrons, yet their modification in nuclear matter remains largely
unconstrained. We report the first derivation of resummed QCD power corrections
to deeply virtual Compton scattering on nuclei. Extending techniques from
inclusive deep inelastic scattering, we identify the nuclear-enhanced
higher-twist contributions generated by coherent final-state scattering of the
struck quark in the medium and resum them to all orders. The corrections are
enhanced by an effective nuclear size $A_{\rm eff}^{1/3}$ and result in an exclusive analogue of dynamical nuclear shadowing. The shift is
controlled by a parameter already fixed by inclusive nuclear data, so no new
nonperturbative input enters. We present quantitative predictions for nuclear
modifications of beam-spin observables at the Electron--Ion
Collider.%
}

\FullConference{The 33rd International Workshop on Deep Inelastic Scattering and Related Subjects (DIS2026)\\
4 - 8 May 2026\\
Bologna, Italy\\}

\begin{document}

\maketitle

\section{Introduction}
GPDs correlate the longitudinal momenta
of partons with their transverse spatial information, and are a principal target of the Electron--Ion
Collider~\cite{Accardi:2012qut,Diehl:2003ny}. Deeply virtual Compton scattering
(DVCS) is the golden channel for accessing them, through a rigorous
leading-twist factorization theorem~\cite{Ji:1996nm} and a
interference with the Bethe--Heitler (BH) process, and two decades of
proton measurements have followed~\cite{Defurne:2017paw}. How
nuclear matter modifies these distributions, by contrast, has received far less
attention, with the first light-nucleus studies appearing only
recently~\cite{Martinez-Fernandez:2026web,Martinez-Fernandez:2026zog}.

The inclusive analogue is well understood: cold nuclear matter distorts the
struck-parton spectrum through intrinsic transverse motion, coherent
re-scattering, and energy loss, effects captured phenomenologically by
nuclear-modified TMDs~\cite{Alrashed:2021csd} and, more recently,
from first principles in soft-collinear effective theory with Glauber
gluons~\cite{Ovanesyan:2011xy,Rothstein:2016bsq,Ke:2023ixa,Ke:2024ytw}. The
exclusive case however obeys a different power
counting. Namely, the cross section falls exponentially in
$t$, so $|t|\sim\LQCD^2$ and the infrared physics is carried by a single
collinear mode, with no soft or transverse-momentum sector for the medium to
modify. Nuclear effects instead enter as coherent higher-twist corrections,
suppressed by $\LQCD^2/Q^2$ but enhanced by powers of the nuclear size. In this
contribution we summarize their resummation. Coherent
final-state scattering of the struck quark generates nuclear-enhanced higher-twist
terms that resum to a shift of the longitudinal momentum argument of the GPD,
Eq.~(\ref{eq: shift}) --- the exclusive analogue of nuclear shadowing in
DIS~\cite{Qiu:2003vd,Vitev:2006bi}, inheriting its single nonperturbative
parameter from the inclusive fits.

\section{Resummed nuclear power corrections}
\label{sec: nuclear}

\subsection{Coherence and the nuclear state}

Extending the vacuum analysis to a nuclear target requires specifying how the
struck quark is embedded in the many-body state. Decomposing the nucleus into
nucleon states and organizing the nuclear wave function by the number of
partonically correlated nucleons gives
\begin{equation}
\ket{A(P)} = \ket{A(P)}_{\rm iso}
           + \sum_{a<b}\ket{A(P)}_{(ab)}
           + \sum_{a<b<c}\ket{A(P)}_{(abc)}
           + \cdots ,
\label{eq: cluster}
\end{equation}
where the first term describes configurations in which the struck quark is
correlated only within a single nucleon, the second a correlated two-nucleon
cluster, and so on up to the nucleus as a whole.

This organization matters because the bilocal quark operator in the Compton
tensor does not act on a single bound nucleon. The struck quark is displaced by
the separation between the two current insertions, and the amplitude survives
only so long as that displacement remains within partonically correlated matter;
translating beyond the correlation region would leave a colored, energetically
disfavored configuration and is strongly suppressed. In the vacuum this region
is the nucleon itself. In a nucleus several nucleons may be partonically
correlated, so the operator acts coherently across a cluster whose size is not
fixed by the nucleon radius but is an emergent property of the nuclear state, $R_{\rm eff} \sim \Aeff^{1/3}/\LQCD$ , $1 \le \Aeff \le A$, interpolating between an uncorrelated nucleus, with no nuclear enhancement at
all, and a fully correlated one. The remaining spectator nucleons source the
color field through which the struck quark propagates and re-scatters; we treat
their collective effect in a mean-field approximation, absorbing it into an
effective pomeron field strength $\mathcal{P}$. Because the strength of the
resulting modification is controlled by the product $\mathcal{P}\,\Aeff^{1/3}$,
a measurement of nuclear shadowing in DVCS probes directly the effective
partonic correlation length of the nuclear state---neither a single bound
nucleon nor the nucleus as a whole---and hence how strongly nucleons are
partonically correlated.

\subsection{Power counting in a nuclear target}

We distinguish the incoherent and coherent channels,
$e+A\to e+N+(A-1)+\gamma^*$ and $e+A\to e+A+\gamma^*$. In the incoherent case
the hard scattering resolves a single nucleon, the factorization theorem is
unchanged relative to the proton, and the nuclear dependence enters only through
the nuclear matrix elements. We therefore focus on the coherent channel, where
the nuclear size enters the power counting directly. An intact nucleus
introduces the nuclear radius $R_A\sim A^{1/3}/\LQCD$ as an additional length
scale, and since the momentum transfer is conjugate to the spatial extent of the
target, $|\Delta_T|\sim1/R_A$. One intriguing possibility is that interactions
generated away from the light-cone trajectory of the struck quark also feed into
the in-medium dynamics and dress the additional nuclear-size dependence carried
by the transverse momentum transfer; although more speculative, this would be
among the most interesting consequences to explore. Collecting the scales, we
have
\begin{equation}
|\Delta_T| \sim A^{-1/3}\LQCD ,
\qquad
|t| \sim \frac{1}{R_A^{2}} \sim A^{-2/3}\LQCD^{2} ,
\qquad
\xi \sim A^{-4/3} .
\label{eq: nuclear counting}
\end{equation}
Both the transverse and the longitudinal momentum transfer are
thus controlled by the nuclear size, and the cross section falls off in $t$ more
rapidly than for the proton.

It remains to identify which contributions carry the nuclear enhancement. The
gluon insertions surviving at leading power are the transversely polarized
fields $\nb^{\alpha}F_{\alpha\beta}$, which encode the transverse color Lorentz
force exerted on the struck quark as it traverses the medium. Each such
insertion costs a factor $A_\perp^\mu/\nb\cdot P\sim\lambda$, so that $n$
insertions naively enter at $\mathcal{O}(\lambda^n)$. The fields may attach to any of the available
nucleons, and summing over these sources produces a combinatorial factor
absorbed into $\mathcal{P}$. Additionally, the integration along the
trajectory of the struck quark returns a factor of the coherence length
$R_{\rm eff}$ for every re-scattering. Each insertion
therefore contributes parametrically $\lambda\,\Aeff^{1/3}$, the
multiple-scattering series is not suppressed, and it must be resummed to all
orders.

\subsection{Multiple scattering and the effective Feynman rules}

To organize the multiple-scattering series we follow the methodology of
Ref.~\cite{Qiu:2003vd}, expanding the time-ordered product of electromagnetic
currents into a chain of $n$ ordered re-scatterings of the struck quark off the
medium. The $u$-channel follows by crossing and is restored at the level of the
Compton tensor by $x\to-x$, so it suffices to track the $s$-channel. Working in
light-cone gauge $\nb\cdot A=0$, the gauge field is traded for the field
strength through $\nb^\alpha F_{\alpha\beta} = \nb\cdot\partial A_\beta$; the
inverse of $\nb\cdot\partial$ is fixed by the retarded prescription
$\nb\cdot\ell\to\nb\cdot\ell-i\epsilon$, the position-space form of a
future-pointing Wilson line, encoding the fact that the re-scattering is a
final-state effect.

At leading power the Dirac chain alternates between $\slashed{n}$ and
$\slashed{\nb}$ projectors separated by the transverse insertions, and a
nonvanishing trace against the external $\gamma^\mu,\gamma^\nu$ requires an
\emph{even} number of insertions: the amplitude is generated by gluon pairs,
and we have
\begin{equation}
\nb\cdot\slashed{F}(y_{n-i})\,\frac{\slashed{\nb}}{2}\,\nb\cdot\slashed{F}(\tilde y_{n-i})
= -\frac{\slashed{\nb}}{2}
\Big(g_\perp^{\alpha\beta} + i\,\epsilon_\perp^{\alpha\beta}\gamma_5\Big)\,
\nb\cdot F_\alpha(y_{n-i})\, \nb\cdot F_\beta(\tilde y_{n-i}) .
\label{eq: pair}
\end{equation}
The symmetric term is a vector coupling that dresses the unpolarized
distributions $H,E$; the antisymmetric term, carrying $\gamma_5$, is an
axial-vector coupling that dresses the polarized $\widetilde H,\widetilde E$.
Because the quark line carries only $\slashed{n}$ and $\slashed{\nb}$, with all
transverse indices residing on the field strengths, the interaction conserves
chirality: the two structures resum independently and no chiral-odd
(transversity) structure is generated.

\begin{figure}[t]
\centering
\includegraphics[width=\linewidth]{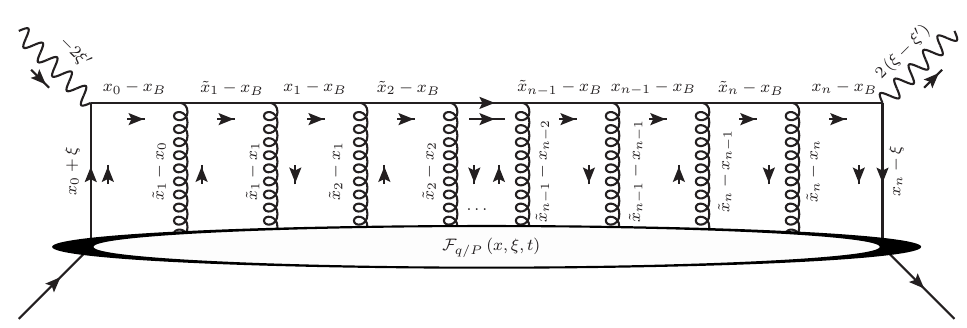}
\caption{Multiple-scattering topology for the $s$-channel Compton tensor and the
effective Feynman rule generated by a pair of insertions.}
\label{fig: rules}
\end{figure}

\subsection{Resummation and the effective momentum shift}

Two ingredients close the series in a compact form. The first is the ordered
integration over the positions of the insertions. Because the interactions are
light-cone ordered, $n\cdot y_{i-1} < n\cdot\tilde y_i < n\cdot y_i$, the nested
integrals collapse to
\begin{equation}
I_n = \int_0^{L_A}\!\! dy_n^-\int_0^{y_n^-}\!\! dy_{n-1}^- \cdots
      \int_0^{y_2^-}\!\! dy_1^-
    = \frac{r_p^{\,n}\,\Aeff^{n/3}}{n!} ,
\label{eq: ordered}
\end{equation}
which supplies both one power of the coherence length per re-scattering, as
anticipated in Sec.~\ref{sec: nuclear}. The second is that each insertion carries a factor of the
conjugate light-cone variable, so that $n$ insertions generate $n$ derivatives
of the distribution with respect to the momentum fraction. The whole effect of coherent multiple scattering is to
evaluate the GPD at a shifted argument,
\begin{equation}
H_q(x,\xi,t) \;\longrightarrow\;
H_q\!\left(x\left[1+\frac{\zeta^{2}\left(A^{1/3}-1\right)}{Q^{2}}\right],\,\xi,\,t\right) ,
\label{eq: shift}
\end{equation}
and likewise for $E$, $\widetilde H$ and $\widetilde E$, which by the argument
below Eq.~(\ref{eq: pair}) resum independently.

The scale $\zeta^2$ appearing here is not a new parameter. In the forward limit
the same manipulation reproduces the known result for nuclear shadowing in
inclusive DIS, in which the structure function is evaluated at the same shifted
momentum fraction, and $\zeta^2 \sim \lim_{x\to0}\tfrac{x}{2}f_{g/N}(x)$
measures the gluon density that the coherent scattering resolves. It was
extracted two decades ago from nuclear DIS data and from open-charm production,
with $\zeta^2\simeq0.09$--$0.12~{\rm GeV}^2$~\cite{Qiu:2003vd,Vitev:2006bi}.
The nuclear DVCS predictions of Sec.~\ref{sec: numerics} therefore contain no
nonperturbative input that has not already been fixed by inclusive measurements.

\section{Numerical results}
\label{sec: numerics}

\subsection{Proton baseline}

Before introducing nuclear effects we must fix the vacuum baseline. We
parameterize the proton GPDs through double distributions, following the
Goloskokov--Kroll construction~\cite{Goloskokov:2006hr}. For a generic
$F\in\{H,E,\widetilde H,\widetilde E\}$ we write
\begin{equation}
F(x,\xi,t) = \int_{-1}^{1}\!\! d\beta \int_{-1+|\beta|}^{1-|\beta|}\!\! d\alpha\;
\delta(x-\beta-\xi\alpha)\, h_F(\beta,t)\, w_n(\beta,\alpha) ,
\qquad
h_F(\beta,t) = h_F(\beta,0)\, e^{\,t\,p_F(\beta)} ,
\label{eq: DD}
\end{equation}
so that the skewness dependence is generated by the Radyushkin profile
$w_n(\beta,\alpha)$ and the $t$ dependence by the Regge-inspired slope
$p_F(\beta) = -\alpha_F'\ln|\beta| + b_F$. Polynomiality is satisfied by
construction, and the forward limits reduce to the unpolarized and polarized
parton distributions, $H_q(x,0,0)=q(x)$ and $\widetilde H_q(x,0,0)=\Delta q(x)$,
which we have verified numerically. The distributions $E$ and $\widetilde E$ are
not constrained by any inclusive forward limit and are modelled by power-law
forward-like ans\"atze; we neglect the $D$ term throughout.

For the observables we use the full BMK cross
section~\cite{Belitsky:2001ns}, retaining the exact kinematic factors
$\mathcal{P}_1$, $\mathcal{P}_2$, $t_{\rm min}$, $K^2$ and the
longitudinal photon flux $\epsilon$, with the BH amplitude built
from the global form-factor parameterization of Ref.~\cite{Ye:2017gyb}. No
parameters are tuned to data at this time.

\begin{figure}[t]
\centering
    \hfill
    \includegraphics[width=0.49\linewidth,valign = c]{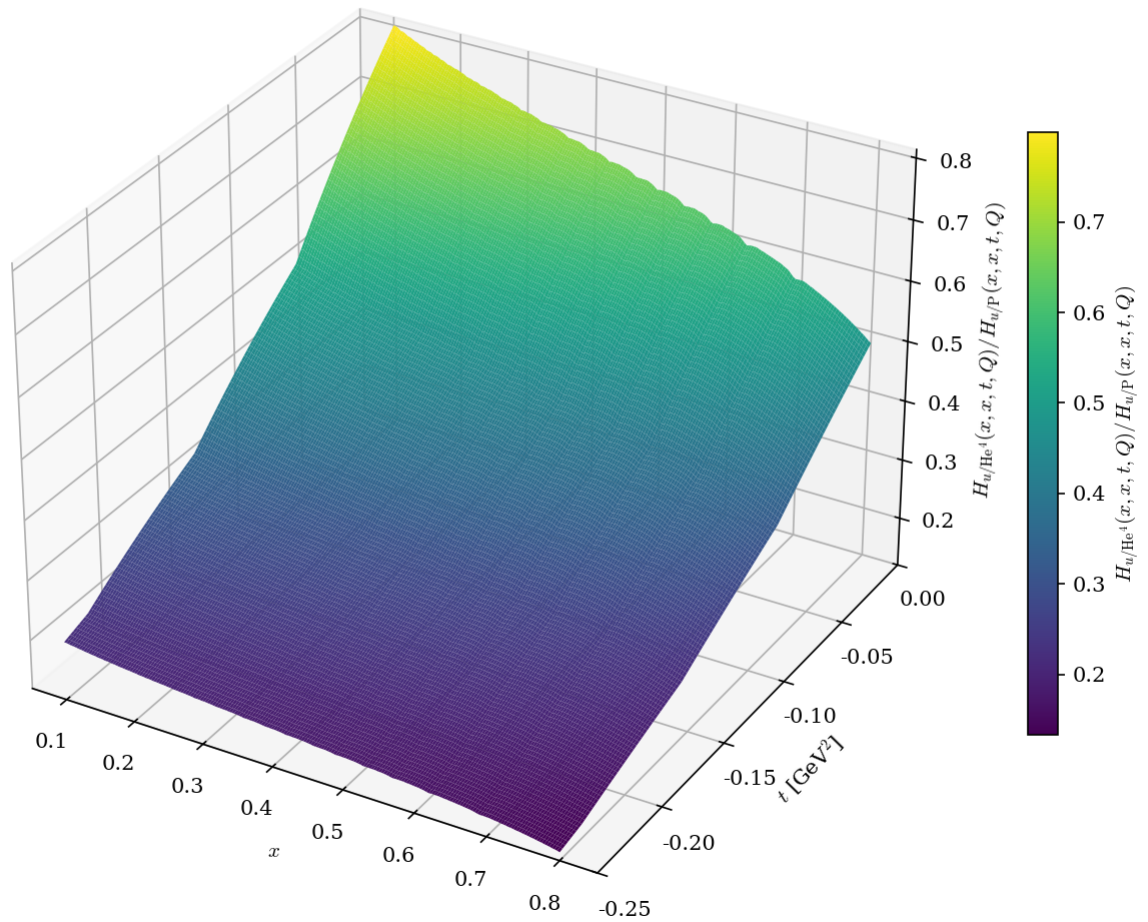}
    \hfill
    \includegraphics[width=0.39\linewidth,valign = c]{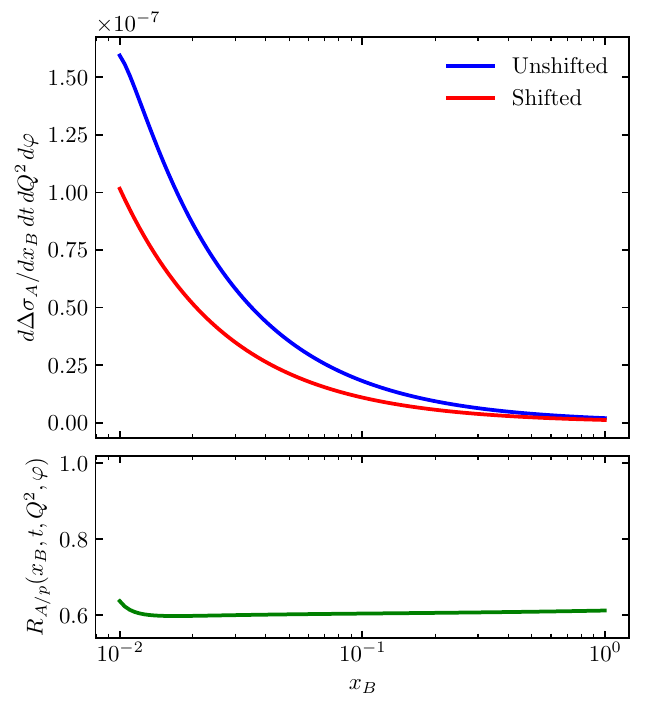}
    \hfill
\caption{\textbf{(a)} Ratio of the up-quark GPD $H$ in $^4$He to that in the
proton on the DVCS diagonal $\xi=x$,
$H_{u/^4{\rm He}}(x,x,t,Q)/H_{u/p}(x,x,t,Q)$, as a function of $x$ and $t$.
\textbf{(b)} Helicity-dependent cross-section difference
$d\Delta\sigma_A/(dx_B\,dt\,dQ^2 d\varphi)$ with and without the medium shift of
Eq.~(\ref{eq: xeff}) (top), and the resulting ratio $R_{A/p}$ of
Eq.~(\ref{eq: ratio}) (bottom).}
\label{fig: predictions}
\end{figure}

\subsection{Nuclear GPDs and predictions}

We now extend the proton parameterization to nuclear targets, restricting
throughout to the coherent process $e+A\to e+A+\gamma$ in which the nucleus
remains intact. The double-distribution machinery of Eq.~(\ref{eq: DD}) is
retained unchanged, the neutron distributions follow from isospin, and the
transverse profile of the proton is replaced by that of the nucleus so as to
avoid double counting,
\begin{equation}
F_A(x,\xi,t) = \frac{F_A^{\rm el}(t)}{F_N^{\rm el}(t)}
\Big[\,Z\,F_p(\xeff,\xi,t) + N\,F_n(\xeff,\xi,t)\,\Big] ,
\qquad
F_A^{\rm el}(t) \simeq \exp\!\big(r_0^{2}A^{2/3}t\big) ,
\label{eq: nuclear ansatz}
\end{equation}
with $r_0\simeq1.2$~fm, so that the nuclear-size dependence of the transverse
momentum transfer enters here only through the static elastic form factor;
whether the coherent dynamics of Sec.~\ref{sec: nuclear} additionally dress this
$A^{2/3}$ scaling, as flagged above, is left to future work. All of the medium
dynamics derived in Sec.~\ref{sec: nuclear} then enter through the single
shifted argument,
\begin{equation}
x_B \;\to\; x_B^{\rm eff} = x_B\left(1+\frac{m_{\rm dyn}^{2}}{Q^{2}}\right) ,
\qquad
m_{\rm dyn}^{2} = \zeta^{2}\big(A^{1/3}-1\big) ,
\label{eq: xeff}
\end{equation}
where $\zeta^{2}\simeq0.09$--$0.12~{\rm GeV}^2$ is the scale controlling
coherent multiple scattering in the medium, fixed long ago from nuclear DIS and
open-charm data~\cite{Qiu:2003vd,Vitev:2006bi} and not refitted here.

Figure~\ref{fig: predictions} shows the resulting helicity-dependent cross
section difference and the ratio
\begin{equation}
R_{A/p}(x_B,t,Q^2,\varphi) =
\frac{d\Delta\sigma_A/(dx_B\,dt\,dQ^2 d\varphi)}
     {A\,d\Delta\sigma_p/(dx_B\,dt\,dQ^2 d\varphi)} ,
\label{eq: ratio}
\end{equation}
for a light nucleus at Jefferson Lab kinematics, $Q^2 = 1~{\rm GeV}^2$ and
$t=-0.25~{\rm GeV}^2$, and for $^{197}{\rm Au}$ at the EIC with
$\sqrt{s} = 29.6~{\rm GeV}$, $Q^2 = 2~{\rm GeV}^2$ and $t=-0.25~{\rm GeV}^2$; in
both cases the event inelasticity is restricted to $0.1<y<0.9$. Since
$m_{\rm dyn}^2>0$, the shift evaluates the GPDs at a larger effective momentum
fraction, where they are falling, and the medium therefore suppresses the
asymmetry. The magnitude of the suppression tracks $m_{\rm dyn}^2/Q^2$: at
Jefferson Lab it is a ten-percent effect over most of the range, steepening
sharply toward large $x_B$, whereas at the EIC the ratio is nearly flat at
$R_{A/p}\approx0.6$, reflecting both the larger $A$ and the smaller $x_B$ reach.

Two caveats should be stated plainly. The curves are central values; we have not
attached uncertainty bands, which would be dominated by $\zeta^2$ and by the
proton GPD input. And we set $\Aeff = A$ throughout, so these represent the
maximal-coherence limit of the picture developed in Sec.~\ref{sec: nuclear}; a
measured suppression weaker than predicted would determine $\Aeff$ directly.

\section{Conclusion}
We have summarized the first systematic treatment of nuclear-enhanced power
corrections in deeply virtual Compton scattering. Starting from the Compton
tensor of the doubly virtual process, we identified coherent final-state
interactions of the struck quark with the nuclear medium as the source of the
dominant higher-twist contributions, and resummed the resulting
multiple-scattering series to all orders. The outcome is an effective shift of
the longitudinal momentum argument of the nuclear GPDs, the exclusive
counterpart of nuclear shadowing in inclusive DIS, requiring no nonperturbative
input beyond the parameter already fixed by inclusive nuclear data.

The predicted suppression of the beam-spin observable reaches the $30$--$50\%$
level in kinematics accessible at the EIC, providing a
perturbative QCD baseline against which future extractions of nuclear GPDs can
be compared. Because its magnitude is set by the effective partonic correlation
length of the nuclear state, such measurements probe directly how strongly
nucleons in a nucleus are partonically correlated.

{\bf Acknowledgments  }
I thank S. Bhattacharya, W. Ke and I. Vitev for their collaboration on this work, and am particularly grateful to I. Vitev for delivering the talk at DIS2026. This work was supported by the U.S. Department of Energy, Office of Science, Office of Nuclear Physics, under contract No. DE-AC02-06CH11357.


\begin{thebibliography}{99}

\bibitem{Accardi:2012qut}
A.~Accardi \textit{et al.},
Eur. Phys. J. A \textbf{52}, no.9, 268 (2016)
[arXiv:1212.1701 [nucl-ex]].

\bibitem{Diehl:2003ny}
M.~Diehl,
Phys. Rept. \textbf{388}, 41-277 (2003)
[arXiv:hep-ph/0307382 [hep-ph]].

\bibitem{Ji:1996nm}
X.~D.~Ji,
Phys. Rev. D \textbf{55}, 7114-7125 (1997)
[arXiv:hep-ph/9609381 [hep-ph]].

\bibitem{Defurne:2017paw}
M.~Defurne \textit{et al.},
Nature Commun. \textbf{8}, no.1, 1408 (2017)
[arXiv:1703.09442 [hep-ex]].

\bibitem{Martinez-Fernandez:2026web}
V.~Martinez-Fernandez, B.~Pire, P.~Sznajder and J.~Wagner,
[arXiv:2604.25677 [hep-ph]].

\bibitem{Martinez-Fernandez:2026zog}
V.~Martinez-Fernandez, B.~Pire, P.~Sznajder and J.~Wagner,
[arXiv:2605.18519 [hep-ph]].

\bibitem{Alrashed:2021csd}
M.~Alrashed, D.~Anderle, Z.~B.~Kang, J.~Terry and H.~Xing,
Phys. Rev. Lett. \textbf{129}, no.24, 242001 (2022)
[arXiv:2107.12401 [hep-ph]].

\bibitem{Fang:2023thw}
S.~Fang, W.~Ke, D.~Y.~Shao and J.~Terry,
JHEP \textbf{05}, 066 (2024)
[arXiv:2311.02150 [hep-ph]].

\bibitem{Ovanesyan:2011xy}
G.~Ovanesyan and I.~Vitev,
JHEP \textbf{06}, 080 (2011)
[arXiv:1103.1074 [hep-ph]].

\bibitem{Rothstein:2016bsq}
I.~Z.~Rothstein and I.~W.~Stewart,
JHEP \textbf{08}, 025 (2016)
[arXiv:1601.04695 [hep-ph]].

\bibitem{Ke:2023ixa}
W.~Ke and I.~Vitev,
Phys. Lett. B \textbf{854}, 138751 (2024)
[arXiv:2301.11940 [hep-ph]].

\bibitem{Ke:2024ytw}
W.~Ke, J.~Terry and I.~Vitev,
JHEP \textbf{02}, 102 (2025)
[arXiv:2408.10310 [hep-ph]].

\bibitem{Schoenleber:2024ihr}
J.~Schoenleber and R.~Szafron,
JHEP \textbf{11}, 031 (2024)
[arXiv:2407.09263 [hep-ph]].

\bibitem{Bhattacharya:2026xxx}
S.~Bhattacharya, W.~Ke, J.~Terry and I.~Vitev,
``Resummed power corrections in deeply virtual Compton scattering,''
in preparation.

\bibitem{Qiu:2003vd}
J.~W.~Qiu and I.~Vitev,
Phys. Rev. Lett. \textbf{93}, 262301 (2004)
[arXiv:hep-ph/0309094 [hep-ph]].

\bibitem{Vitev:2006bi}
I.~Vitev, J.~T.~Goldman, M.~B.~Johnson and J.~W.~Qiu,
Phys. Rev. D \textbf{74}, 054010 (2006)
[arXiv:hep-ph/0605200 [hep-ph]].

\bibitem{Belitsky:2001ns}
A.~V.~Belitsky, D.~Mueller and A.~Kirchner,
Nucl. Phys. B \textbf{629}, 323-392 (2002)
[arXiv:hep-ph/0112108 [hep-ph]].

\bibitem{Goloskokov:2006hr}
S.~V.~Goloskokov and P.~Kroll,
Eur. Phys. J. C \textbf{50}, 829-842 (2007)
[arXiv:hep-ph/0611290 [hep-ph]].

\bibitem{Ye:2017gyb}
Z.~Ye, J.~Arrington, R.~J.~Hill and G.~Lee,
Phys. Lett. B \textbf{777}, 8-15 (2018)
[arXiv:1707.09063 [nucl-ex]].

\end{thebibliography}
\end{document}